\documentclass{ws-ijgmmp}
\usepackage{amsmath}   
\usepackage{amssymb}   
\usepackage{bm}
\usepackage{graphicx}
\usepackage{fix-cm}

\pdfoutput=1

\usepackage{array}
\usepackage{color}
\newcommand\ii{\'{\i}}

\def\Dsl{\,\raise.15ex\hbox{/}\mkern-13.5mu D} 

\usepackage{hyperref}

\usepackage{epstopdf}

\newcommand{\beq}{\begin{eqnarray}}
\newcommand{\eeq}{\end{eqnarray}}
\newcommand{\R}{\mathbb{R}}

\newcommand{\eref}[1]{(\ref{#1})}
\newcommand{\nsl}{\vec{\gamma}\cdot\vec{n}\,}
\def\Dsl{\,\raise.15ex\hbox{/}\mkern-12.5mu D}
\def\dsl{\,\raise.15ex\hbox{/}\mkern-12.5mu \partial}

\begin{document}

\markboth{M. Asorey}
{Boundary Effects in Field Theories}
\def\draftnote{}

\def\ii{\'{\i}}

%
\catchline{}{}{}{}{}
%

\title{Boundary Effects in Bosonic and Fermionic Field Theories
}

\author{M. Asorey}

\address{ Departamento de F\ii sica Te\'orica, Facultad de Ciencias, Universidad de Zaragoza\\
E-50009 Zaragoza, Spain\\
asorey@unizar.es}

\author{D. Garc\ii a-Alvarez}

\address{Departamento de An\'alisis Econ\'omico \\ Facultad de Econom\ii a  y  Empresa.\\
 Universidad de Zaragoza. E-50005 Zaragoza. Spain\\
dga@unizar.es}

\author{J. M.  Mu\~noz-Casta\~neda}

 \address{Institut f\"ur Theoretische Physik, Universit\"at Leipzig, \\
Br\"uderstr. 16, D-04103 Leipzig, Germany\\
 jose.munoz-castaneda@uni-leipzig.de}

\maketitle

\begin{history}
\received{(Day Month Year)}
\revised{(Day Month Year)}
\end{history}

\begin{abstract}
The dynamics of quantum field theories  on bounded domains 
requires the introduction of  boundary conditions on the quantum fields.
We address the problem from a very general perspective by using charge conservation
as a fundamental principle for scalar and fermionic quantum field theories. 
Unitarity arises as a consequence of the choice of charge preserving
boundary conditions.
This provides a powerful framework for the analysis of  global geometrical  
and topological properties of the space of
 physical boundary conditions.
Boundary conditions which allow the existence of edge states can only arise in theories 
with a mass gap which is also a physical requirement for topological
insulators.
\end{abstract}


\section{Introduction}	

Since the early days of the quantum theory boundary effects arise in quantum physics. 
In  Young's double slit experiments, which provide many key ingredients
of the foundations of the quantum theory, boundary effects play a crucial role.
Boundary effects also appear as relevant ingredients in
the Aharonov-Bohm effect, which points out the quantum observability
of phase factors of electromagnetic fields.

More recently, a plethora  of new quantum effects induced by the presence of boundaries
boosted a new era of quantum technologies. Some of  the most remarkable new phenomena include
the Casimir effect, the presence of plasmons and other surface effects
in metals and dielectrics, the appearance of edge currents in the Hall effect, the opening of a gap
in small graphene samples and new edge effects in topological insulators.

From a more basic viewpoint boundary effects also
appear in fundamental physics:  black hole horizons effects, Hawking radiation,
topological defects, topology change and holographic effects in the AdS/CFT correspondence.

The increasing relevance of boundary effects is demanding  a comprehensive theory of
boundary conditions.
In spite of the fact that quite a lot of work has been devoted to establish the foundations
of the quantum theory, a comprehensive theory of boundary conditions for quantum theories
is still missing. A first attempt to fill the gap was initiated by Asorey-Ibort-Marmo in Ref.
\cite{aim} and was further developed in \cite{amc}. In this paper we will try to emphasize some
new aspects of this approach in a relativistic context and illustrate the emerging general theory by means of
examples.

 In quantum mechanics  the fundamental  principle of the theory of boundary conditions  is the preservation of 
 probability. Indeed, unitarity imposes severe constraints on the boundary behavior of quantum states in systems confined to bounded domains \cite{aim}. However, in relativistic field theories, the fundamental principle is charge conservation, which together with  causality imposes further conditions \cite{amc}. 
 The space of boundary conditions compatible with both constraints  has interesting global geometric properties.
The dependence of many  interesting physical phenomena, like the Casimir effect \cite{adj1},
topology change \cite{bala,aim,wilczek, survey}  or  renormalization group flows \cite{adj2}, on the boundary conditions can be  analyzed from this global perspective.

\section{Boundary effects in Quantum Mechanics}
To illustrate the large variety  of boundary conditions which are compatible with the foundations
of  quantum mechanics let us consider a free particle of unit mass confined in an  one dimensional interval $[0,L]$.
Even in classical mechanics the dynamics of a free particle is not completely defined by the Hamiltonian
$$ H=\frac{p^2}{2}.$$
Once the particle reach the boundaries at $x=0$ or $x=L$ one needs to specify how  the particle 
bounces back into the interval. There are several possibilities, which depend on the physical properties of 
the boundaries: complete reflection, sticky reflection or even a complete stop
of the particle motion  at the boundary (see \cite{gift}).
In quantum mechanics the situation is similar. The naive quantization rule gives rise to
a Hamiltonian 
$$ H= -\frac12 {d^2\over
dx^2}$$
which is not selfadjoint operator in $L^2([0,L])$ and, thus, does not univocally determines the quantum dynamics.

Even if $H$ is symmetric for smooth functions of compact support there are many other functions
on $L^2([0,L])$ where $H$ is not symmetric. For instance  for two arbitrary smooth functions on $[0,L]$ we have
$$(\psi_1,H\psi_2)=-\frac12\int_0^L \psi_1^\ast
\psi_2'' \, dx.$$
By integrating by parts one obtains 
$$(\psi_1,H\psi_2)-(\psi_2,H\psi_1)=-\frac12\int_0^L  {d\over
dx}[\psi_1^\ast
\psi_2'-{\psi_1'}^\ast \psi_2]=\Sigma(\psi_1,\psi_2).$$
The obstruction to the symmetry of $H$ arises from the boundary term
$$\Sigma(\psi_1,\psi_2)=\psi_1^\ast
\psi_2'(L)-{\psi_1'}^\ast
\psi_2(L)-\psi_1^\ast
\psi_2'(0)-{\psi_1'}^\ast\psi_2(0).$$

 This boundary term can only vanish on a dense subset of functions on $L^2([0,L]$
 if there is an $2\times 2$ unitary matrix   
\beq
U=\left(\begin{array}{cc}u_{11} & u_{12}  \\u_{11} & u_{22} 
 \end{array}\right)
\eeq
such that the boundary values of the functions in the domain of the Hamiltonian satisfy 
the following boundary conditions \cite{aim,Tsu}
$$\left(\begin{array}{cc}\varphi(0)-i\dot\varphi(0)\\
\varphi(L)-i\dot\varphi(L) \end{array}\right)=\left(\begin{array}{cc}u_{11} &u_{12}\\
u_{21}&u_{22} \end{array}\right)\left(\begin{array}{cc}\varphi(0)+i\dot\varphi(0)\\
\varphi(L)+i\dot\varphi(L) \end{array}\right),$$
where $\dot\varphi(0)=-\psi'(0)$ and $\dot\varphi(L)=\psi'(L)$.
In other terms, the definition of quantum dynamics has many different implementations
parametrized by the different selfadjoint extension of $H$, which are in one to one correspondence
with the unitary matrices of $U(2)$ \cite{aim,Tsu}.

Some specially interesting examples correspond to the case when the
matrix $U$ is diagonal or anti-diagonal. In the first case we have
\beq
U_\alpha=\left(\begin{array}{cc}e^{-i\alpha_1}&0\\
0&e^{-i\alpha_2}\end{array}\right)
\label{Robin}
\eeq
which corresponds to the Robin boundary conditions
\beq
-\sin
{\alpha_1\over 2}\varphi(0)+\cos{\alpha_1\over 2}\dot\varphi(0)=0\nonumber\\ 
-\sin {\alpha_2\over 2}\varphi(L)+\cos{\alpha_2\over 2}\dot\varphi(L)=0\nonumber
 \eeq
which includes Newmann $\dot\varphi(0)=\dot\varphi(L)=0$ and
Dirichlet $\varphi(0)=\varphi(L)=0$ boundary conditions.
	In the anti-diagonal case  
$$U_\epsilon=\left(\begin{array}{cc}0&e^{-i\epsilon}\\
e^{i\epsilon}&0\end{array}\right)
$$
we have pseudo-periodic boundary conditions 
$\varphi(L)=e^{i\epsilon}\varphi(0)$ 
with a probability flux propagating from one boundary to the other. The $U_0=\sigma_1$ matrix with $\epsilon=0$
corresponds to periodic boundary conditions $\varphi(0)=\varphi(L)$ and that with $\epsilon=\pi$
to anti-periodic boundary conditions  $\varphi(0)=-\varphi(L)$ .

Another interesting non-diagonal case is described by the unitary matrix
\beq
U_g= \frac1{1-ig} 
\begin{pmatrix}
i g&{1}\cr
{1}&{i g} 
\end{pmatrix},
\label{delta}
\eeq
\begin{figure}[hb]
\centering
  \includegraphics[width=3cm]{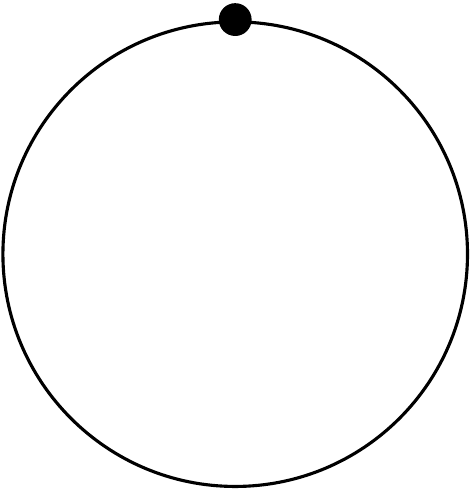}
  \caption{The dynamics of a  superconducting quantum device (SQUID) can be described by the
  boundary conditions introduced by the matrix $U_g$  of equation \eref{delta}
   } \label{topchange}
\end{figure}
which corresponds to a delta like potential on a point of a circle \cite{Tsu,Tsu2,Tsu3}. 
This boundary condition describes the effective dynamics  of circular superconducting quantum device (SQUID)  with
a Josephson junction (see Figure 1). In this case the non-diagonal form of the 
boundary condition is accounting for the tunneling effects from one side 
of the interval to the other, which in this particular case corresponds to a delta like potential  in the
circle \cite{Tsu,Tsu2}. All other non-diagonal boundary conditions correspond to more general contact interactions
combined with magnetic fluxes crossing the circle.

The generalization for several intervals is straightforward \cite{bala,aim}. 
In the case of two intervals $[-2L,-L]\cup [L,2L]$ the boundary conditions are given by 
$$\begin{array}{cc} \end{array}\left(\begin{array}{cc}\varphi(-2)-i\dot\varphi(-2)\\
\varphi(-L)-i\dot\varphi(-L) \\
\varphi(L)-i\dot\varphi(L)\\
\varphi(2L)-i\dot\varphi(2L)
 \end{array}\right)=\left(\begin{array}{cccc}
u_{11} &u_{12}&u_{13}&u_{14}\\
u_{21}&u_{22}&u_{23}&u_{24}\\
u_{31}&u_{32}&u_{33}&u_{34}\\
u_{41}&u_{42}&u_{43}&u_{44}\\
 \end{array}\right)\left(\begin{array}{cc}\varphi(-2)+i\dot\varphi(-2)\\
\varphi(-L)+i\dot\varphi(-L) \\
\varphi(L)+i\dot\varphi(L)\\
\varphi(2L)+i\dot\varphi(2L)\\
\end{array}\right),$$
in terms of a four dimensional unitary matrix  $U\in U(4)$, where $\dot\varphi(-2L)=-\psi'(-2L)$,  $\dot\varphi(-L)=\psi'(-L)$
$\dot\varphi(L)=-\psi'(L)$ and $\dot\varphi(2L)=\psi'(2L)$.
The non-diagonal elements of the boundary conditions describe boundary conditions where the global conservation
of the probability flux is obtained by cancellation of the non-trivial magnetic fluxes between the four different edges 
of the  intervals. For instance, the unitary matrix
\beq
U_1=\begin{pmatrix}
{0}&{0}&{0}&{1}\cr
{0}&{0}&{1}&{0}\cr
{0}&{1}&{0}&{0}\cr
{1}&{0}&{0}&{0}\cr
\end{pmatrix}\quad
\eeq
corresponds to periodic boundary conditions between the  edges $-L$ and $L$ and $2L$ and $-2L$  which 
describe the quantum dynamics on circle of length 2, whereas the unitary matrix
\beq
U_2=\begin{pmatrix}
{0}&{1}&{0}&{0}\cr
{1}&{0}&{0}&{0}\cr
{0}&{0}&{0}&{1}\cr
{0}&{0}&{1}&{0}\cr
\end{pmatrix}
\eeq
corresponds to periodic boundary conditions between the  edges $-2L$ and $-L$, and $L$ and $2L$  which 
describe the quantum dynamics on two independent circles of unit length. Thus, the transition from a double connected 
topology to a single connected one described in Figure 2 can be carried out in a smooth way by changing the boundary 
conditions  along the  one-parameter family  $U_s=U_1^s U_2^{1-s}, s\in[0,1]$  of unitary matrices of  $U(4)$, interpolating between the matrices $U_2$ and $U_1$.  In this way a topology fluctuation phenomena can be described in very a simple manner in terms of boundary conditions \cite{aim,wilczek,survey}.
\vspace{-.4cm}
\begin{figure}[ht]
\centering
  \includegraphics[width=10cm]{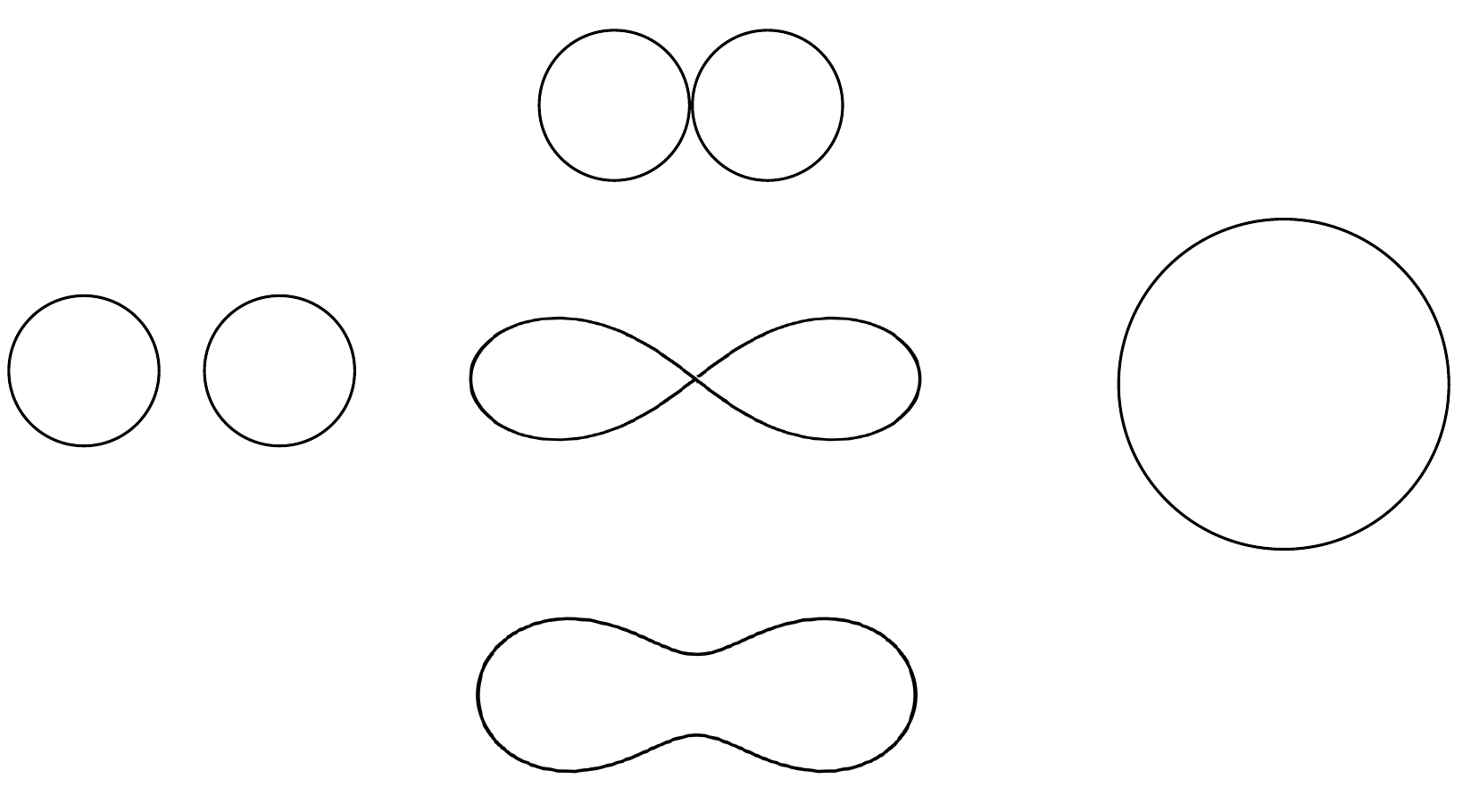}
  \caption{Topology change induced by a continuous change of boundary conditions $U_s=U_1^s U_2^{1-s}, s\in[0,1]$. The transition from a two circle topology $U_2$ to one circle topology $U_1$ is made through a contact point  topology $U_\infty$.
   } \label{topchange}

\end{figure}

\begin{figure}[b]
\centering
  \includegraphics[width=5cm]{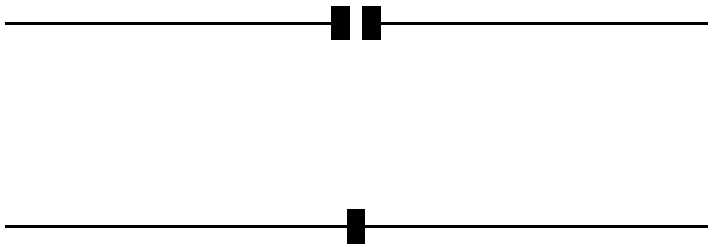}
  \caption{ The effect of a singular boundary between two complementary  domains  can be described by boundary conditions on the 
  inner/outer interface boundary  between the two domains.
   } \label{topchange}
\end{figure}

The absolute principle of strict conservation of the quantum  probability  in a bounded domain $\Omega\subset\mathbb{R}$ is not realistic, because most of the physical boundaries
are not completely insulating. 
The most general boundary conditions should account for more realistic boundaries, where
a net probability  flux can cross the boundary and the absolute law of probability conservation only holds in the whole space. 
This can be achieved by doubling the boundary $\partial\Omega$ and considering two systems: an inner
system in $\Omega$ and an outer system in $\mathbb{R}\backslash \Omega$, both sharing a common boundary $\partial\Omega$.

Let us illustrate the method with a simple example. 
Let us consider, for simplicity the splitting of the real
line into two unbounded semi-intervals $(-\infty,-L]$ and $[L,\infty)$ (see Figure 3). It is easy to check that the 
only boundary conditions that cancel  probability flux in the whole system are again given  by 
$$\left(\begin{array}{cc}\varphi(-L)-i\dot\varphi(-L)\\
\varphi(L)-i\dot\varphi(L) \end{array}\right)=\left(\begin{array}{cc}u_{11} &u_{12}\\
u_{21}&u_{22} \end{array}\right)\left(\begin{array}{cc}\varphi(-L)+i\dot\varphi(-L)\\
\varphi(L)+i\dot\varphi(L) \end{array}\right),$$
in terms of a $2\times 2$
unitary matrix $U\in U(2)$. The boundary condition  \eref{delta}  given by the matrix $U_g$  
in this case  corresponds to a $-g\delta(x)$ potential
sitting in the origin (as $L\to 0$) of the real line $\mathbb{R}$, i.e.
\beq
\varphi(L)=\varphi(-L)\hspace{.8cm}\nonumber\\
\varphi'(L)-\varphi'(-L)=-2 g \varphi(L)\nonumber§.
\eeq

In the case of an attractive potential $-g\delta(x)$ with $g>0$, there is a negative energy bound state of
the form $\psi(x)=\textstyle\sqrt{g}\, e^{-{g} |x|}$\cite{mgadella, mcmg1,mcmg2},
which corresponds to a physical state highly localized at the impurity, i.e. an edge state. 
The same state also appears as a negative mode for the Robin boundary conditions with a diagonal unitary matrix 
of type \eref{Robin} with $\tan\frac{\alpha_1}{2}=\tan\frac{\alpha_1}{2}=-{g}$ and $\pi<\alpha_1=\alpha_2< 2 \pi$.
This is general feature than can happen for
boundary conditions whose unitary matrices have an eigenvalue $e^{i\alpha}$ with $\pi<\alpha< 2\pi $ \cite{aim, amc} which provides natural candidates of suitable boundary conditions for topological insulators with edge states \cite{tpi, bala2}.

\section{Boundary conditions and charge conservation}

We have shown that unitarity is the fundamental quantum principle of the theory of boundary conditions
in non-relativistic quantum mechanics. However, boundary conditions also appears in many other
classical and quantum  physical systems governed by 
partial differential equations as  necessary conditions to uniquely determine the  dynamical evolution
of the physical system when is confined on a bounded domain.

In  the theory of boundary conditions in quantum mechanics there is  another equivalent principle,
which  can be easily  generalized  and allows the extension of the formalism beyond the
framework of non-relativistic quantum mechanics. Indeed, we have seen  that unitarity is related to probability
conservation, but this is also is related to charge conservation, which ultimately is a consequence the existence of a 
$U(1)$ internal symmetry. 
Indeed, the conservation of  the $U(1)$ symmetry can be used 
as the fundamental principle in  the theory of physical boundary conditions.
In the presence of a $U(1)$ symmetry  Noether's theorem 
implies the existence of a conserved current
$$\partial^\mu j_\mu=0.$$
Thus, the dynamical variation  of the charge $\rho=j_0$ in a given subdomain $\Omega$ 
can be expressed in terms of the net flux  of the
current $\vec{j}$ across its boundary boundary $\partial \Omega$,i.e.
$$ \frac{d}{dt}{\int_\Omega \rho\, dv}=\int_{\partial\Omega} \vec{j}\cdot \,d\vec{\sigma}.$$
A good example is non-relativistic quantum mechanics, where the Schr\"odinger equation
$i\partial_t \Psi=-\frac12\nabla^2 \psi +V(x)\psi$
preserves the $U(1)$ symmetry
$$\psi'(x)=e^{i\alpha}\psi(x).$$
The corresponding conserved current
$$\vec{j} = \frac{i}{2}[\psi^\ast \vec{\nabla} \psi-(\vec{\nabla} \psi^\ast) \psi]$$
encodes the probability conservation law.

This example also illustrate how two quantum symmetries can be related beyond  simple
  compatibility.
In this case  unitarity of time evolution, or which is equivalent, time translation invariance,  is 
equivalent to the existence of $U(1)$ charge symmetry. In other terms, 
charge conservation is strongly related to the self-adjointness of the Hamiltonian operator $H$ governing the time
evolution.
Indeed, the probability conservation law implies that
$$\partial_t \langle\Psi(t)|\psi(t)\rangle=  \langle \partial_t\Psi(t)|\psi(t)\rangle+ \langle\Psi(t)| \partial_t\psi(t)\rangle=
i \langle\Psi(t) |(H^\dagger- H) \psi(t)\rangle=0,
$$
i.e. the self-adjointness  of the Hamiltonian operator $H$, $H^\dagger=H$. Conversely, self-adjointness of the
Hamiltonian operator implies the unitary of time evolution and, thus, probability conservation.  
In this case both principles 
are equivalent. To better understand the crucial role played by semi-bounded quadratic forms  
in the theory of boundary conditions see Ref. \cite{lledo}.

\section{Boundary conditions in Scalar Field Theories}

In a complex scalar field theory which preserves the electric charge, i.e. the $U(1)$ gauge symmetry 
$$\phi'(x)=e^{i\alpha}\phi(x),$$
gives rise to the conserved current  
$$j_\mu = \textstyle  \frac{i}{2}[\phi^\ast \partial_\mu \psi-(\partial_\mu \phi^\ast) \phi].$$
Notice that in this case the charge density 
\beq
\rho=j_0 = \textstyle  \frac{i}{2}[\phi^\ast \partial_t \psi-(\partial_t \phi^\ast) \phi]
\label{cdkg}
\eeq
is not definite positive.
Thus, the connection with unitary time evolution is not so evident. First of all, there is not
a positive Hilbert product to associate  with an unitary evolution. However, charge conservation
can still be  a good fundamental principle to define consistent boundary conditions when the  system is
confined in a bounded  spatial domain $\Omega\subset \R^3$.  Physically consistent
boundary conditions  must  enforce 
the vanishing of the charge flux across the boundary $\partial\Omega$ of the system. 
In fact, since the 
spatial current is the same as in non-relativistic quantum mechanics, the theory of 
boundary conditions is also the same. Indeed, the boundary term accounting for the charge flux
across the boundary is
$$\Sigma(\phi)=\int_{\partial\Omega} \vec{j}\cdot d\vec{\sigma}= \frac{i}{2} \int_{\partial\Omega}
[\phi^\ast \vec{\nabla} \phi-(\vec{\nabla} \phi^\ast) \phi] d\vec{\sigma}.
$$
If we now consider  a linear combination $\phi=\phi_1+\alpha \phi_2$ of two independent fields $\phi_1,\phi_2$,
 such that $\Sigma(\phi_1)=\Sigma(\phi_2)=0$, we obtain
$$\Sigma(\phi)=   \frac{i \alpha}{2} \int_{\partial\Omega}
[\phi_1^\ast \vec{\nabla} \phi_2-(\vec{\nabla} \phi_1^\ast) \phi_2] d\vec{\sigma}+ \frac{i \alpha^\ast}{2} \int_{\partial\Omega}
[\phi_2^\ast \vec{\nabla} \phi_1-(\vec{\nabla} \phi_2^\ast) \phi_1] d\vec{\sigma}
$$
which implies that 
$$\int_{\partial\Omega}
[\phi_1^\ast \vec{\nabla} \phi_2-(\vec{\nabla} \phi_1^\ast) \phi_2] d\vec{\sigma}=\int_{\partial\Omega}
[\phi_2^\ast \vec{\nabla} \phi_1-(\vec{\nabla} \phi_2^\ast) \phi_1] d\vec{\sigma}=0,
$$
and if denote by $\partial_n=\vec{\nabla}\cdot \vec{n}$ the normal derivative at the boundary $\partial \Omega$,
$$\int_{\partial\Omega}
[\phi_1^\ast\partial_n \phi_2-(\partial_n \phi_1^\ast) \phi_2] d{\sigma}=0.
$$
Thus, the most general boundary condition without singular zero-modes, 
which preserves the charge density conservation law, is given  by
\beq
(1- i\partial_n)\phi=U(1+ i\partial_n)\phi
\label{bckg}
\eeq
in terms of  a unitary
matrix $U$ defined on the boundary Hilbert space $L^2(\Omega)$ space.
Notice that the above boundary conditions guarantee that the spacial Laplacian operator $-\Delta_U$ is selfadjoint with
respect to the standard product $(\cdot,\cdot)$ of $L^2(\Omega)$ \cite{aim}.
In fact, all these selfadjoint extensions of $-\Delta$ can be defined by  \eref{bckg} in terms of unitary matrices of the Hilbert space
$L^2(\Omega)$ \cite{gadella,  amc}.

Let us consider a boundary
condition such that $-\Delta_U$ is a positive operator.
Then, we can define  a  pseudo-Hilbert product in the subspace  ${\cal D}(\sqrt{-\Delta_U})\oplus{\cal D}(\sqrt{-\Delta_U})$ of Hilbert space ${\cal H}= {\cal H}_+\oplus{\cal H}_-$, with $
{\cal H}_+={\cal H}_-=L^{2}(\Omega)$,
given by 
$$(\phi,\phi)_H=(\phi_+,\sqrt{-\Delta_U}\phi_+)-(\phi_-,\sqrt{-\Delta_U}\phi_-)$$
for any $\phi_+,\phi_- $in the domain  ${\cal D}(\sqrt{-\Delta_U})$ of $\sqrt{-\Delta_U}$ with   $\phi=\phi_+ +\phi_-\in {\cal H}.$
Notice that the Hamiltonian of the associated free field theory $H=\pm\sqrt{-\Delta_U}$ is selfadjoint with respect to
the $(\cdot,\cdot)_H$ product, 
\beq
(\phi,H\phi)_H=  (\phi_+,H^2 \phi_+)+(\phi_-,H^2\phi_-).
\label{hpos}
\eeq
But not any  selfadjoint extension of $H^2$ corresponds to one of $H$ with respect to the product
$(\cdot,\cdot)_H$, only the positive ones satisfy this property. Finally, we remark
 that the pseudo hermitian product $(\cdot,\cdot)_H$ is nothing but the product associated to the conserved
 charge density \eref{cdkg}.
 
 An interesting remark is that the Hamiltonian of the free field theory is selfadjoint with respect to both products $(\cdot,\cdot)$ and $(\cdot,\cdot)_H$. 
 There is, however, a significative difference between both cases. $H$ is
not definite positive or negative
 with respect to the standard product $(\cdot,\cdot)$ of ${\cal H}_+\oplus{\cal H}_-$ 
 $$ (\phi,H\phi)=  (\phi_+,H \phi_+)+(\phi_-,H\phi_-)= (\phi_+,\sqrt{-\Delta_U} \phi_+)-(\phi_-,\sqrt{-\Delta_U}\phi_-),$$
whereas with respect to the $H$-product it is selfadjoint and positive \eref{hpos}.

Finally, we remark that although we have emphasized the role of charge conservation in the definition of selfadjoint
Hamiltonian for the free field theory, there are extra requirements which further constrain the range of physically 
consistent boundary conditions. In particular, as it was already mentioned, unitarity of the quantum field theory
also requires that the operator $-\Delta_U$ has to be positive, to make possible the self-adjointness of the restriction
 of the Hamiltonian of the  quantum field theory to the one-particle states $\sqrt{-\Delta_U}$ \cite{amc}.  This means that only boundary conditions defined by unitary matrices
with eigenvalues $e^{i\alpha}$ with $0\leq \alpha\leq \pi$ are physically consistent for any size of the physical space $\Omega$
\cite{amc}.

\section{Boundary conditions in Fermionic Field Theories}
In a fermionic Dirac field theory which preserves the electric charge,  the $U(1)$ gauge symmetry 
$$\psi'(x)=e^{i\alpha}\psi(x),$$
defines the conserved current is 
$$j_\mu =   \overline{\psi}\gamma_\mu \psi.$$
Notice that in this case, unlike the bosonic case, the charge density is definite positive
\beq
\rho=j_0 =   \overline{\psi}\gamma_0 \psi=\psi^\dagger\psi.
\label{cdkg}
\eeq
  Charge conservation
becomes again a  fundamental principle to define consistent boundary conditions.
Physically consistent boundary conditions should imply the 
vanishing of the charge current flux 
$$\vec{j}=  \overline{\psi}\vec{\gamma} \psi=\psi^\dagger \gamma_0 \vec{\gamma} \psi$$
across the boundary of the system $\partial\Omega$. 
The boundary term accounting for the charge flux
across the boundary is
$$\Sigma(\psi)=\int_{\partial\Omega} \vec{j}\cdot d\vec{\sigma}=  \int_{\partial\Omega}
\psi^\dagger \gamma_0\vec{\nabla} \psi\, d\vec{\sigma} = \int_{\partial\Omega}
\psi^\dagger \gamma_0\nsl \psi\, d{\sigma} .
$$

Again, as in the bosonic case, if we consider  a linear combination $\psi=\psi_1+\alpha \psi_2$ of two independent fields $\psi_1,\psi_2$,
 such that $\Sigma(\psi_1)=\Sigma(\psi_2)=0$, we obtain
$$\Sigma(\psi)=   { \alpha} \int_{\partial\Omega}
\psi_1^\dagger \gamma_0\nsl \psi_2\, d{\sigma}+ { \alpha^\ast}\int_{\partial\Omega}
\psi_2^\dagger \gamma_0\nsl \psi_1\, d{\sigma}
$$
which implies that the vanishing condition of the charge flux of $\psi$ is
$$\int_{\partial\Omega}
\psi_1^\dagger \gamma_0\nsl \psi_2\, d{\sigma}=\int_{\partial\Omega}
\psi_2^\dagger \gamma_0\nsl \psi_1\, d{\sigma}
=0,
$$
The boundary term can be split as the difference of  two positive chiral components 

$$
\int_{\partial\Omega}
\psi_1^\dagger \gamma_0\nsl \psi_2\, d{\sigma}= \int_{\partial\Omega}{\psi_1^+}^\dagger \psi_2^+ \, d{\sigma}-\int_{\partial\Omega}
{\psi_1^-}^\dagger  \psi_2^-\, d{\sigma}
$$
where ${\psi^\pm}=(1\pm \gamma_0\nsl)\psi$.

  Thus, the most general boundary condition preserving charge conservation  is given by ${\psi^-=U \gamma_{0} \psi^+}$, i.e.
 \begin{equation}\label{cdos}
(1-\gamma_0\nsl)\psi=U \gamma_{0} (1+\gamma_0\nsl)\psi,
 \end{equation}
 where $U$ is any unitary operator of the Hilbert space of boundary spinors, which 
   anti-commute with $\gamma_0\nsl$,
 i.e. $\{U,\gamma_0\nsl\}=0$ \cite{bala}. 

The striking feature is that even if we select conservation of charge as the fundamental fundamental principle to
fix the boundary conditions of the theory, we get as a bonus  that these boundary conditions guarantee that
the Hamiltonian of the free field theory
$H=i\gamma^0\vec{\gamma}\cdot\vec{\nabla}-m\gamma^0$
  is selfadjoint with respect to the product defined by the charge form
$$(\psi_1,\psi_2)= \int_{\Omega}{\psi_1}^\dagger \psi_2\, dv$$ 
i.e.
 $$(\psi_1,H\psi_2)=(H\psi_1,\psi_2).$$
This property together with time independence of the Dirac Hamiltonian ($\partial_t H=0$) also  implies
time translation invariance of the free field theory
\beq
\partial_t(\psi, H \psi)&=&(\partial_t \psi^\dagger, H \psi)+ ( \psi^\dagger,(\partial_t H) \psi)+
( \psi^\dagger, H \partial_t \psi)\\
&=& i (H \psi^\dagger, H \psi)-i (\psi^\dagger, H^2 \psi)=i(\psi^\dagger, (H^\dagger-H)H \psi)=0.
\eeq
By construction the Hilbert product $(\cdot,\cdot) $ is time translation invariant but the boundary conditions
imply that this property is also preserved by the dynamics of the system, which  guarantees   the self-adjointness of
the free field Hamiltonian, even if in this case it is not a positive operator.

In consequence, even if we give up the choice of unitarity as the fundamental principle to fix the boundary conditions
of the system, we finally have a selfconsistent approach where charge conservation is the driving fundamental principle
and unitarity is also recovered as a byproduct.

\section*{Acknowledgements}

Many of the results summarized in this paper were obtained in papers writen in collaboration with many other authors  to whom we would like to thank: N. Acharya,  A.P. Balachandran, J. Clemente-Gallardo, A. Ibort, G. Marmo, J.M. P\'erez-Prado, A. Queiroz and  S. Vaidya. M. A.  work has been partially supported by the Spanish MICINN grants  FPA2012-35453 and CPAN Consolider Project CDS2007-42 and DGA-FSE (grant 2014-E24/2).

\end{document}